*Review*
# A Brief History of Gravitational Waves

**Jorge L. Cervantes-Cota [1], Salvador Galindo-Uribarri [1] and George F. Smoot [2,3,4,]***

[1] Department of Physics, National Institute for Nuclear Research, Km 36.5 Carretera Mexico-Toluca, Ocoyoacac, Mexico State C.P.52750, Mexico; jorge.cervantes@inin.gob.mx (J.L.C.-C.); salvador.galindo@inin.gob.mx (S.G.-U.)

[2] Helmut and Ana Pao Sohmen Professor at Large, Institute for Advanced Study, Hong Kong University of Science and Technology, Clear Water Bay, 999077 Kowloon, Hong Kong, China.

[3] Université Sorbonne Paris Cité, Laboratoire APC-PCCP, Université Paris Diderot, 10 rue Alice Domon et Leonie Duquet 75205 Paris Cedex 13, France.

[4] Department of Physics and LBNL, University of California; MS Bldg 50-5505 LBNL, 1 Cyclotron Road Berkeley, CA 94720, USA.

* Correspondence: gfsmoot@lbl.gov; Tel.:+1-510-486-5505

**Abstract:** This review describes the discovery of gravitational waves. We recount the journey of predicting and finding those waves, since its beginning in the early twentieth century, their prediction by Einstein in 1916, theoretical and experimental blunders, efforts towards their detection, and finally the subsequent successful discovery.

1. Introduction

Einstein's General Theory of Relativity, published in November 1915, led to the prediction of the existence of gravitational waves that would be so faint and their interaction with matter so weak that Einstein himself wondered if they could ever be discovered. Even if they were detectable, Einstein also wondered if they would ever be useful enough for use in science. However, exactly 100 years after his theory was born, on 14 September 2015, these waves were finally detected and are going to provide scientific results.

In fact at 11:50:45 am CET on 14 September 2015 Marco Drago—a postdoc—was seated in front of a computer monitor at the Max Planck Institute for Gravitational Physics in Hanover, Germany, when he received an e-mail, automatically generated three minutes before from the monitors of LIGO (for its acronym Laser Interferometer Gravitational wave Observatory). Marco opened the e-mail, which contained two links. He opened both links and each contained a graph of a signal similar to that recorded by ornithologists to register the songs of birds. One graph came from a LIGO station located at Hanford, in Washington State, and the other from Livingston Station in the state of Louisiana [1].

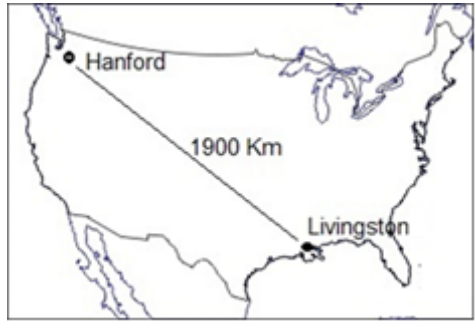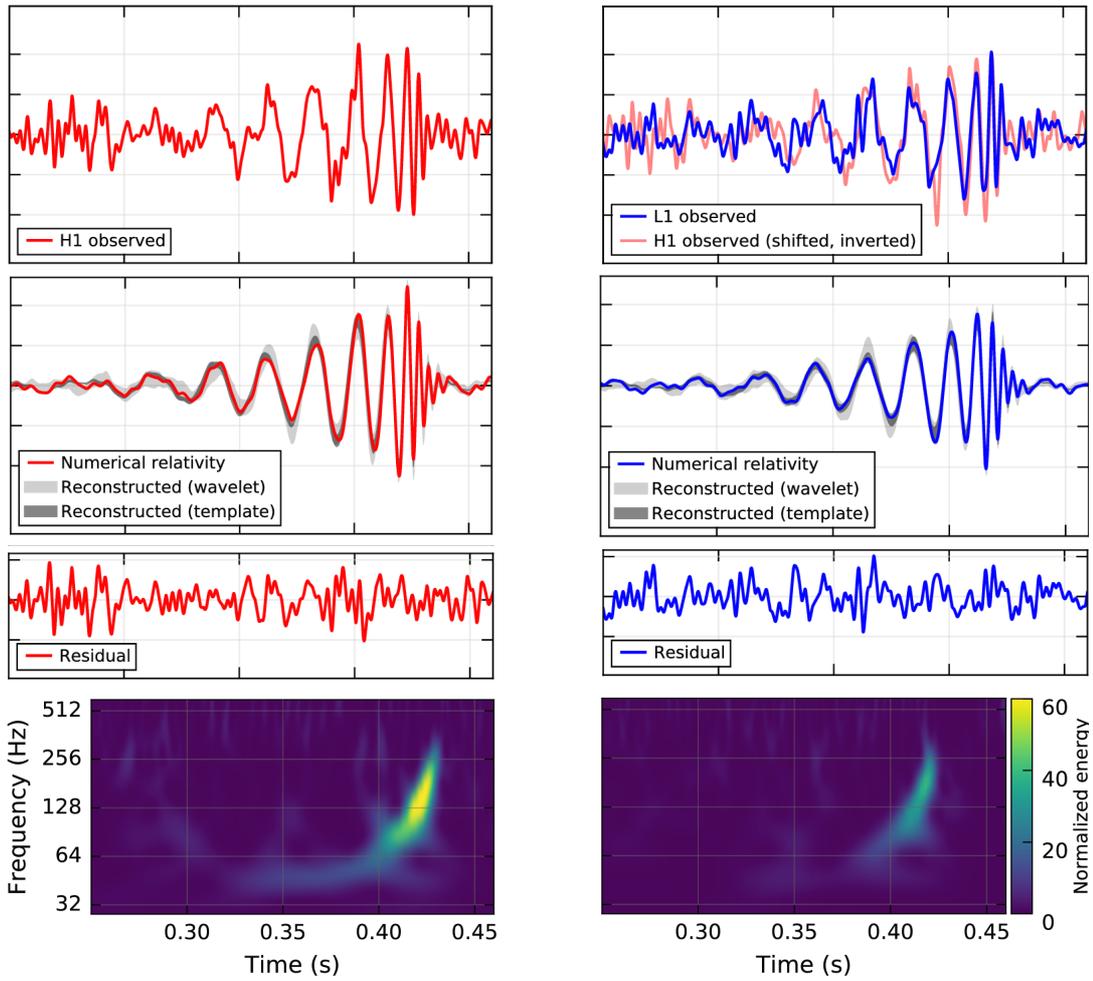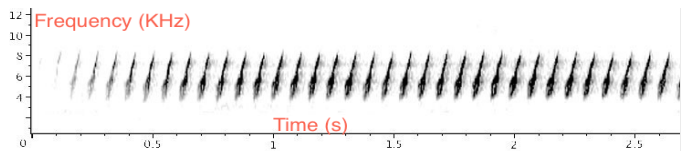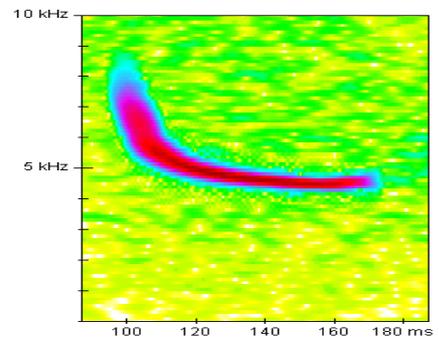

(a)

(b)

(c)

**Figure 1.** (**a**): Waveforms from LIGO sites [2] and their location and sonograms. Figure from https://losc.ligo.org/events/GW150914/. The gravitational-wave event GW150914 observed by the LIGO Hanford (H1, left column panels) and Livingston (L1, right column panels) detectors. Times are shown relative to 14 September 2015 at 09:50:45 UTC. For visualization, all time series are filtered with a 35–350 Hz band-pass filter to suppress large fluctuations outside the detectors' most sensitive frequency band, and band-reject filters to remove the strong instrumental spectral lines. (**b**): Chirping Sparrow (Spizella passerina) song: frequency versus time (in seconds), showing a song made up of a series of chirps. (**c**): A Pipistrelle bat call for echolocation. Bats use ultrasound for "seeing" and for social calls. This spectrogram is a graphic representation of frequencies against time. The color represents the loudness of each frequency. This spectrogram shows a falling call, which becomes a steady note. The yellow and green blotches are noise. This is nearly a time reverse of black hole merger sonograms.

Marco is a member of a team of 30 physicists working in Hanover, analyzing data from Hanford and Livingston. Marco's duty is to be aware of and analyze the occurrence of an "event" that records the passage of a gravitational wave, in one of the four lines that automatically track the signals from the detectors in the two LIGO observatories on the other side of the Atlantic.

Marco noticed that the two graphs were almost identical, despite having been registered independently in sites separated by 1900 km (see Figure 1a); for comparison we include sonograms from animals (see Figures 1b and 1c). The time that elapsed between the two signals differed by about 7 milliseconds. These almost simultaneous records of signals coming from sites far away from each other, the similarity of their shapes and their large size, could not be anything but, either: a possible record of a gravitational wave traveling at the speed of light or, a "signal" artificially "injected" to the detectors by one of the four members of the LIGO program who are allowed to "inject" dummy signals. The reason why artificial signals are injected to the system is to the test whether the operation of the detectors is correct and if the duty observers are able to identify a real signal.

Following the pre-established protocols, Marco tried to verify whether the signals were real or the "event" was just a dummy injected signal. Since aLIGO was still in engineering mode, there was no way to inject fake signals, i.e., hardware injections. Therefore, everyone was nearly 100% certain that this was a detection. However, it was necessary to go through the protocols of making sure that this was the case.

Marco asked Andrew Lundgren, another postdoc at Hanover, to find out if the latter was the case. Andrew found no evidence of a "dummy injection." On the other hand, the two signals detected were so clear, they did not need to be filtered to remove background noise. They were obvious. Marco and Andrew immediately phoned the control rooms at Livingston and Hanford. It was early morning in the United States and only someone from Livingston responded. There was nothing unusual to report. Finally, one hour after receiving the signal, Marco sent an e-mail to all collaborators of LIGO asking if anyone was aware of something that might cause a spurious signal. No one answered the e-mail.

Days later, LIGO leaders sent a report stating that there had not been any "artificial injection." By then the news had already been leaked to some other members of the world community of astrophysicists. Finally after several months, the official news of the detection of gravitational waves was given at a press conference on 11 February 2016, after the team had ascertained that the signals were not the result of some experimental failure, or any signal locally produced, earthquake, or electromagnetic fluctuation. This announcement is the most important scientific news so far this century. Gravitational waves were detected after 60 years of searching and 100 years since the prediction of their existence. The scientific paper was published in *Physical Review Letters* [2]. This discovery not only confirmed one of the most basic predictions of General Relativity but also opened a new window of observation of the universe, and we affirm without exaggeration that a new era in astronomy has been born.

In what follows we shall narrate the journey experienced in search of gravitational waves, including their conception in the early twentieth century, their prediction by Einstein in 1916, the theoretical controversy, efforts towards detection, and the recent discovery.

**2. Lost and Found Gravitational Waves**

On 5 July 1905 the *Comptes Rendus of the French Academy of Sciences* published an article written by Henri Poincare entitled "Sur la dynamique d' l'électron." This work summarized his theory of relativity [3]. The work proposed that gravity was transmitted through a wave that Poincaré called a gravitational wave (*onde gravifique*).

It would take some years for Albert Einstein to postulate in 1915 in final form the Theory of General Relativity [4]. His theory can be seen as an extension of the Special Theory of Relativity postulated by him 10 years earlier in 1905 [5]. The General Theory explains the phenomenon of gravity. In this theory, gravity is not a force—a difference from Newton's Law—but a manifestation of the curvature of space–time, this curvature being caused by the presence of mass (and also energy and momentum of an object). In other words, Einstein's equations match, on the one side, the curvature of local space–time with, on the other side of the equation, local energy and momentum within that space–time.

Einstein's equations are too complicated to be solved in full generality and only a few very specific solutions that describe space–times with very restrictive conditions of symmetry are known. Only with such restrictions it is possible to simplify Einstein's equations and so find exact analytical solutions. For other cases one must make some simplifications or approximations that allow a solution, or there are cases where equations can be solved numerically using computers, with advanced techniques in the field called Numerical Relativity.

Shortly after having finished his theory Einstein conjectured, just as Poincaré had done, that there could be gravitational waves similar to electromagnetic waves. The latter are produced by accelerations of electric charges. In the electromagnetic case, what is commonly found is dipolar radiation produced by swinging an electric dipole. An electric dipole is formed by two (positive and negative) charges that are separated by some distance. Oscillations of the dipole separation generate electromagnetic waves. However, in the gravitational case, the analogy breaks down because there is no equivalent to a negative electric charge. There are no negative masses. In principle, the expectation of theoretically emulate gravitational waves similar to electromagnetic ones faded in Einstein's view. This we know from a letter he wrote to his colleague, Karl Schwarzschild, on 19 February 1916. In this letter, Einstein mentioned in passing:

*"Since then [November 14] I have handled Newton's case differently, of course, according to the final theory [the theory of General Relativity]. Thus there are no gravitational waves analogous to light waves. This probably is also related to the one-sidedness of the sign of the scalar T, incidentally [this implies the nonexistence of a "gravitational dipole"] [6].*

However, Einstein was not entirely convinced of the non-existence of gravitational waves; for a few months after having completed the General Theory, he refocused efforts to manipulate his equations to obtain an equation that looked like the wave equation of electrodynamics (Maxwell's wave equation), which predicts the existence of electromagnetic waves. However, as mentioned, these equations are complex and Einstein had to make several approximations and assumptions to transform them into something similar to Maxwell's equation. For some months his efforts were futile. The reason was that he used a coordinate system that hindered his calculations. When, at the suggestion of a colleague, he changed coordinate systems, he found a solution that predicted three different kinds of gravitational waves. These three kinds of waves were baptized by Hermann Weyl as longitudinal-longitudinal, transverse-longitudinal, and transverse-transverse [7].

These approaches made by Einstein were long open to criticism from several researchers and even Einstein had doubts. In this case Einstein had manipulated his field equations into a first approximation for wave-emitting bodies whose own gravitational field is negligible and with waves that propagate in empty and flat space.

Yet the question of the existence of these gravitational waves dogged Einstein and other notable figures in the field of relativity for decades to come. By 1922 Arthur Eddington wrote an article entitled "The propagation of gravitational waves" [8]. In this paper, Eddington showed that two of the three types of waves found by Einstein could travel at any speed and this speed depends on the coordinate system; therefore, they actually were spurious waves. The problem Eddington found in Einstein's original calculations is that the coordinate system he used was in itself a "wavy" system and therefore two of the three wave types were simply flat space seen from a wavy coordinates system; i.e., mathematical artifacts were produced by the coordinate system and were not really waves at all. So the existence of the third wave (the transverse-transverse), allegedly traveling at the speed of light, was also questioned. Importantly, Eddington did prove that this last wave type propagates at the speed of light in all coordinate systems, so he did not rule out its existence.

In 1933 Einstein emigrated to the United States, where he had a professorship at the Institute for Advanced Study in Princeton. Among other projects, he continued to work on gravitational waves with the young American student Nathan Rosen.

In 1936 Einstein wrote to his friend, renowned physicist Max Born, "Together with a young collaborator [Rosen], I arrive at the interesting result that *gravitational waves do not exist*, though they have been assumed a certainty to the first approximation" (emphasis added) [9] (p. 121, Letter 71).

That same year, Einstein and Rosen sent on 1 June an article entitled "Are there any gravitational waves?" to the prestigious journal *Physical Review*, whose editor was John T. Tate [10]. Although the original version of the manuscript does not exist today, it follows from the abovementioned letter to Max Born that the answer to the title of the article was "they do not exist."

The editor of the *Physical Review* sent the manuscript to Howard Percy Robertson, who carefully examined it and made several negative comments. John Tate in turn wrote to Einstein on 23 July, asking him to respond to the reviewer's comments. Einstein's reaction was anger and indignation; he sent the following note to Tate [10]:

*July 27, 1936*
*Dear Sir.*
*"We (Mr. Rosen and I) had sent you our manuscript for publication and had not authorized you to show it to specialists before it is printed. I see no reason to address the—in any case erroneous—comments of your anonymous expert. On the basis of this incident I prefer to publish the paper elsewhere."*
*Respectfully*
*Einstein*
*P.S. Mr. Rosen, who has left for the Soviet Union, has authorized me to represent him in this matter.*

On July 30th, John Tate replied to Einstein that he very much regretted the withdrawal of the article, saying "I could not accept for publication in *The Physical Review* a paper which the author was unwilling I should show to our Editorial Board before publication" [10].

During the summer of 1936 a young physicist named Leopold Infeld replaced Nathan Rosen as the new assistant to Einstein. Rosen had departed a few days before for the Soviet Union. Once he arrived at Princeton, Infeld befriended Robertson (the referee of the Einstein–Rosen article). During one of their encounters the topic of gravitational waves arose. Robertson confessed to Infeld his skepticism about the results obtained by Einstein. Infeld and Robertson discussed the point and reviewed together the Einstein and Rosen manuscript, confirming the error. Infeld in turn informed Einstein about the conversation with Robertson.

An anecdote illustrating the confused situation prevailing at that time is given in Infeld's autobiography. Infeld refers to the day before a scheduled talk that Einstein was to give at Princeton on the "Nonexistence of gravitational waves." Einstein was already aware of the error in his manuscript, which was previously pointed out by Infeld. There was no time to cancel the talk. The next day Einstein gave his talk and concluded, *"If you ask me whether there are gravitational waves or not, I must answer that I don't know. But it is a highly interesting problem"* [10].

After having withdrawn the Einstein–Rosen paper from the *Physical Review*, Einstein had summited the very same manuscript to the *Journal of the Franklin Society* (Philadelphia). This journal accepted the paper for publication without modifications. However, after Einstein learned that the paper he had written with Rosen was wrong, he had to modify the galley proofs of the paper. Einstein sent a letter to the editor on 13 November 1936 explaining the reasons why he had to make fundamental changes to the galley proofs. Einstein also renamed the paper, entitling it "On gravitational waves," and modified it to include different conclusions [10]. It should be noted that this would not have happened if Einstein had accepted in the first instance Robertson's valid criticisms. Tellingly, the new conclusions of his rewritten article read [11]:

"Rigorous solution for Gravitational cylindrical waves is provided. For convenience of the reader the theory of gravitational waves and their production, known in principle, is presented in the first part of this article. After finding relationships that cast doubt on the existence of gravitational fields rigorous wavelike solutions, we have thoroughly investigated the case of cylindrical gravitational waves. As a result, there are strict solutions and the problem is reduced to conventional cylindrical waves in Euclidean space."

Furthermore, Einstein included this explanatory note at the end of his paper [11],

*"Note—The second part of this article was considerably altered by me after the departure to Russia of Mr. Rosen as we had misinterpreted the results of our formula. I want to thank my colleague Professor Robertson for their friendly help in clarifying the original error. I also thank Mr. Hoffmann your kind assistance in translation."*

In the end, Einstein became convinced of the existence of gravitational waves, whereas Nathan Rosen always thought that they were just a formal mathematical construct with no real physical meaning.

### 3. Pirani's Trip to Poland; the Effect of a Gravitational Wave

To prove the existence of a gravitational wave it becomes necessary to detect its effects. One of the difficulties presented by the General Theory of Relativity resides in how to choose the appropriate coordinate system in which one observer may calculate an experimentally measurable quantity, which could, in turn, be compared to a real observation. Coordinate systems commonly used in past calculations were chosen for reasons of mathematical simplification and not for reasons of physical convenience. In practice, a real observer in each measurement uses a local Cartesian coordinate system relative to its state of motion and local time. To remedy this situation, in 1956 Felix A. E. Pirani published a work that became a classic article in the further development of the Theory of Relativity. The article title was "On the physical significance of the Riemann tensor" [12]. The intention of this work was to demonstrate a mathematical formalism for the deduction of physical observable quantities applicable to gravitational waves. Curiously, the work was published in a Polish magazine. The reason for this was that Pirani, who at that time worked in Ireland, went to Poland to visit his colleague Leopold Infeld, of whom we have already spoken; the latter had returned to his native Poland in 1950 to help boost devastated Polish postwar physics. Because Infeld went back to Poland, and because of the anti-communist climate of that era, Infeld was stripped of his Canadian citizenship. In solidarity Pirani visited Poland and sent his aforementioned manuscript to *Acta Physica Polonica*. The importance of Pirani's Polish paper is that he used a very practical approach that got around this whole problem of the coordinate system, and he showed that the waves would move particles back and forth as they pass by.

## 4. Back and Forth as Waves Pass by

One of the most famous of Einstein's collaborators, Peter Bergmann, wrote a well-known popular book *The Riddle of Gravitation*, which describes the effect a gravitational wave passing over a set of particles would have [13]. Following Bergmann, we shall explain this effect.

When a gravitational wave passes through a set of particles positioned in an imaginary circle and initially at rest, the passing wave will move these particles. This motion is perpendicular (transverse) to the direction in which the gravitational wave travels. For example, suppose a gravitational pulse passes in a direction perpendicular to this page, Figure 2 shows how a set of particles, initially arranged in a circle, would sequentially move (a, b, c, d).

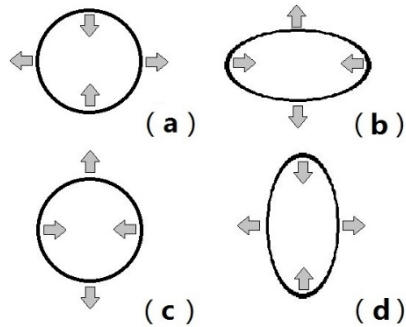

**Figure 2.** (**a**,**b**,**c**,**d**) Sequential effect of a gravitational wave on a ring of particles. In the image of Figure 2**a** is observed as the particles near horizontal move away from each other while those are near vertical move together to reach finally the next moment as shown in (Figure 2**b**). At that moment all the motions are reversed and so on. This is shown in Figure 2**c**,**d**. All these motions occur successively in the plane perpendicular (transverse) to the direction of wave propagation.

At first sight the detection of gravitational waves now seems very simple. One has to compare distances between perpendicularly placed pairs of particles and wait until a gravitational wave transits. However, one has to understand in detail how things happen. For instance, a ruler will not stretch, in response to a gravitational wave, in the same way as a free pair of particles, due to the elastic properties of the ruler, c.f. note 11, p. 19 in [14]. Later it was realized that changes produced in such disposition of particle pairs (or bodies) can be measured if, instead of the distances, we measure the time taken by light to traverse them, as the speed of light is constant and unaltered by a gravitational wave.

Anyway, Pirani's 1956 work remained unknown among most physicists because scientists were focusing their attention on whether or not gravitational waves carry energy. This misperception stems from the rather subtle matter of defining energy in General Relativity. Whereas in Special Relativity energy is conserved, in General Relativity energy conservation is not simple to visualize. In physics, a conservation law of any quantity is the result of an underlying symmetry. For example, linear momentum is conserved if there is spatial translational symmetry, that is, if the system under consideration is moved by a certain amount and nothing changes. In the same way, energy is conserved if the system is invariant under time. In General Relativity, time is part of the coordinate system, and normally it depends on the position. Therefore, globally, energy is not conserved. However, any curved space–time can be considered to be locally flat and, locally, energy is conserved.

During the mid-1950s the question of whether or not gravitational waves would transmit energy was still a hot issue. In addition, the controversy could not be solved since there were no experimental observations that would settle this matter. However, this situation was finally clarified thanks to the already mentioned work by Pirani [12], and the comments suggested by Richard Feynman together with a hypothetical experiment he proposed. The experiment was suggested and comments were delivered by Feynman during a milestone Congress held in 1957 in Chapel Hill,

North Carolina. We will come back to this experiment later, but first we shall speak about the genesis of the Chapel Hill meeting.

**5. What Goes Up Must Come Down**

The interest in the search for gravitational waves began at a meeting occurred in Chapel Hill, North Carolina in 1957. The meeting brought together many scientists interested in the study of gravity. What is unusual is that this meeting would not have been possible without the funding of an eccentric American millionaire named Roger W. Babson.

On 19 January 1949 Roger W. Babson founded the Gravity Research Foundation (GRF), which still exists today. Babson's motivation for establishing the foundation was a "debt" that he thought he owed to Newton's laws—which, according to his understanding, led him to become a millionaire [15]. Babson earned the greatest part of his fortune in the New York Stock Exchange by applying his own version of Newton's Gravity law, "What goes up must come down." Thus he bought cheap shares on their upward route and sold them before their price collapsed. His ability to apply the laws of Newton was surprising because he anticipated the 1929 Wall Street crash. "To every action there is a reaction," he used to preach.

Babson's interest in gravity arose when he was a child, following a family tragedy. Babson's older sister drowned when he was still an infant. In his version of the unfortunate accident, he recalls, "... she could not fight gravity." The story of this eccentric millionaire is detailed on the website of the GRF foundation [15]. Babson became obsessed with finding a way to control the force of gravity and therefore he established the aforementioned foundation, which had as its main activity arranging a yearly essay contest that dealt with "the chances of discovering a partial insulation, reflector, or absorber of gravity." An annual award of $1000 (a considerable amount at the time) was offered to the best essay. The essays submitted for the competition were limited to 2000 words. This annual award attracted several bizarre competitors and was awarded several times to risible submissions. However, in 1953 Bryce DeWitt, a young researcher at Lawrence Livermore Laboratory in California, decided to write an essay and enter the contest because he needed the money to pay his home's mortgage.

The essay presented by DeWitt in the 1953 competition was a devastating critique of the belief that it is possible to control gravity. In DeWitt's own words, his writing "essentially nagged [the organizers] for that stupid idea" [16]. To his surprise, his essay was the winner despite having been written in one night. DeWitt notes those were "the faster 1000 dollars earned in my whole life!" [16].

But DeWitt never imagined he would earn many thousands more dollars than he won with his essay. The reason for it might be found in the final paragraph of his essay,

"In the near future, external stimuli to induce young people to engage in gravitational physics research, despite its difficulties, are urgently needed" [16].

This final paragraph of DeWitt's essay echoed in Babson's mind. Perhaps, he thought, why not focus my philanthropy to support serious studies of gravitation? Perhaps he thought his GRF could refocus its activities onto the scientific study of gravitation.

Babson shared this new enthusiasm with a friend, Agnew Bahnson, also a millionaire and also interested in gravity. Bahnson was a little more practical than Babson and convinced him to found an independent institute separate from GRF. Thus arose the idea of founding a new Institute of Field Physics (IOFP), whose purpose would be pure research in the gravitational fields. The idea of founding the IOFP was clever because the old GRF was severely discredited among the scientific community. For example, one of the promotional brochures GRF mentioned as an example of the real possibility of gravity control the biblical episode where Jesus walks on water. Such was the ridicule and vilification of the GRF in scientific circles that a famous popularizer of mathematics, Martin Gardner, devoted an entire chapter of one of his books to ridiculing the GRF. In this work Gardner claims that the GRF "is perhaps the most useless project of the twentieth century" [17]. So Bahnson knew that to research on gravity in the discredited GRF had very little chance of attracting serious scientists. We must mention that today GRF enjoys good prestige and many well-known scientists have submitted their essays to its annual competition. That proves that it is worth trying for a thousand dollars.

In order to start the new IOFP institute off on the right foot, Bahnson contacted a famous Princeton physicist, John Archibald Wheeler, who supported the idea of hiring Bryce DeWitt to lead the new institute, whose headquarters would be established in Chapel Hill, NC, Bahnson's hometown and headquarters of the University of North Carolina. Wheeler, knowing the vast fortune of the couple of millionaires, hastened to send a telegram to DeWitt. In one of his lines the telegram said "Please do not give him a 'no' for answer from the start" [16]. That's how DeWitt won more than one thousand dollars, actually much more. In January 1957 the IOFP was formally inaugurated, holding a scientific conference on the theme "The role of gravitation in Physics." As we shall review below, the Chapel Hill conference rekindled the crestfallen and stagnant study of gravitation prevailing in those days.

## 6. The Chapel Hill Conference 1957

The 1957 Chapel Hill conference was an important event for the study of gravity. Attendance was substantial: around 40 speakers from institutions from 11 countries met for six days, from 18 to 23 January 1957 on the premises of the University of North Carolina at Chapel Hill. Participants who attended the meeting were predominantly young physicists of the new guard: Feynman, Schwinger, Wheeler, and others. During the six-day conference, discussions focused on various topics: classical gravitational fields, the possibility of unification of gravity with quantum theory, cosmology, measurements of radio astronomy, the dynamics of the universe, and gravitational waves [18].

The conference played a central role in the future development of classical and quantum gravity. It should be noted that the Chapel Hill 1957 conference today is known as the GR1 conference. That is, the first of a series of GR meetings that have been held regularly in order to discuss the state of the art in matters of Gravitation and General Relativity (GR = General Relativity). The conference has been held in many countries and possesses international prestige. The last was held in New York City in 2016.

In addition to the issues and debates on the cosmological models and the reality of gravitational waves, during the conference many questions were formulated, including ideas that are topical even today. To mention a few, we can say that one of the assistants, named Hugh Everett, briefly alluded to his parallel universes interpretation of quantum physics. On the other hand, DeWitt himself pointed out the possibility of solving gravitation equations through the use of electronic computers and warned of the difficulties that would be encountered in scheduling them for calculations, thus foreseeing the future development of the field of Numerical Relativity. However, what concerns us here is that gravitational waves were also discussed at the conference; chiefly, the question was whether gravitational waves carrying energy or not.

Hermann Bondi, a distinguished physicist at King's College London, presided over session III of Congress entitled "General Relativity not quantized." In his welcome address to the participants he warned "…still do not know if a transmitter transmits energy radiation …" [18] (p. 95). With these words Bondi marked the theme that several of the speakers dealt with in their presentations and subsequent discussion. Some parts of the debate focused on a technical discussion to answer the question about the effect a gravitational pulse would have on a particle when passing by, i.e., whether or not the wave transmits energy to the particle.

During the discussions, Feynman came up with an argument that convinced most of the audience.

His reasoning is today known as the "sticky bead argument". Feynman's reasoning is based on a thought experiment that can be described briefly as follows: Imagine two rings of beads on a bar (see Figure 3, upper part). The bead rings can slide freely along the bar. If the bar is placed transversely to the propagation of a gravitational wave, the wave will generate tidal forces with respect to the midpoint of the bar. These forces in turn will produce longitudinal compressive stress on the bar. Meanwhile, and because the bead rings can slide on the bar and also in response to the tidal forces, they will slide toward the extreme ends first and then to the center of the bar (Figure 3, bottom). If contact between the beads and the bar is "sticky," then both pieces (beads and bar) will be heated by friction. This heating implies that energy was transmitted to the bar by the gravitational wave, showing that gravitational waves carry energy [18].

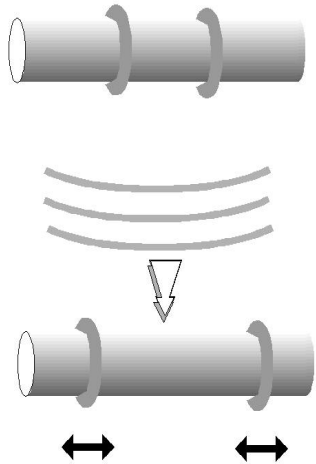

**Figure 3.** Sketch of the "sticky bead argument".

In a letter to Victor Weisskopf, Feynman recalls the 1957 conference in Chapel Hill and says, "I was surprised to find that a whole day of the conference was spent on this issue and that 'experts' were confused. That's what happens when one is considering energy conservation tensors, etc. instead of questioning, can waves do work?" [19].

Discussions on the effects of gravitational waves introduced at Chapel Hill and the "sticky bead argument" convinced many—including Hermann Bondi, who had, ironically, been among the skeptics on the existence of gravitational waves. Shortly after the Chapel Hill meeting Bondi issued a variant of the "sticky bead argument" [20].

Among the Chapel Hill audience, Joseph Weber was present. Weber was an engineer at the University of Maryland. He became fascinated by discussions about gravitational waves and decided to design a device that could detect them. Thus, while discussions among theoretical physicists continued in subsequent years, Weber went even further because, as discussed below, he soon began designing an instrument to make the discovery.

## 7. The First Gravitational Wave Detector

The year following the meeting at Chapel Hill, Joseph Weber began to speculate how he could detect gravitational waves. In 1960 he published a paper describing his ideas on this matter [21]. Basically he proposed the detection of gravitational waves by measuring vibrations induced in a mechanical system. For this purpose, Weber designed and built a large metal cylinder as a sort of "antenna" to observe resonant vibrations induced in this antenna that will eventually be produced by a transit of a gravitational wave pulse. This is something like waiting for someone to hit a bell with a hammer to hear its ring.

It took his team several years to build the "antenna," a task that ended by the mid-sixties. In 1966, Weber, in a paper published in *Physical Review*, released details of his detector and provided evidence of its performance [22]. His "antenna" was a big aluminum cylinder about 66 cm in diameter and 153 cm in length, weighing 3 tons. The cylinder was hanging by a steel wire from a support built to isolate vibrations of its environment (see Figure 4). In addition, the whole arrangement was placed inside a vacuum chamber. To complete his instrument, Weber placed a belt of detectors around the cylinder. The detectors were piezoelectric crystals to sense cylinder vibrations induced by gravitational waves. Piezoelectric sensors convert mechanical vibrations into electrical impulses.

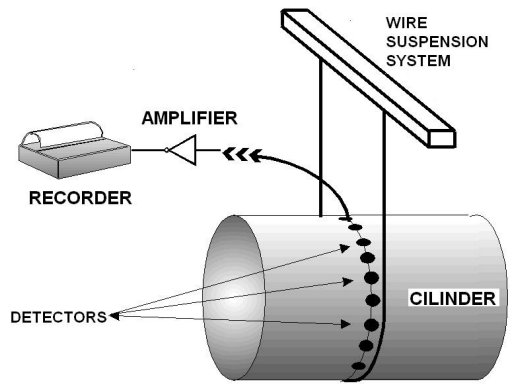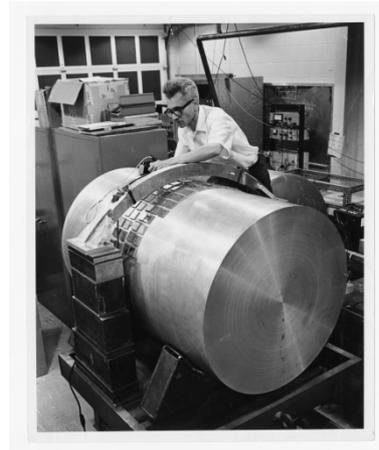

**Figure 4.** Sketch of Weber's cylinder detector and photo of Joseph Weber at the antenna.

Weber built two detectors. The first one was at the University of Maryland and the other was situated 950 km away, in Argonne National Laboratory near Chicago. Both detectors were connected to a registration center by a high-speed phone line. The idea of having two antennas separated by a large distance allowed Weber to eliminate spurious local signals, that is, signals produced by local disturbances such as thunderstorms, cosmic rays showers, power supply fluctuations, etc. In other words, if a detected signal was not recorded simultaneously in both laboratories, the signal should be discarded because it was a local signal and therefore spurious.

For several years, Weber made great efforts to isolate his cylinders from spurious vibrations, local earthquakes, and electromagnetic interference, and argued that the only significant source of background noise was random thermal motions of the atoms of the aluminum cylinder. This thermal agitation caused the cylinder length to vary erratically by about $10^{-16}$ meters, less than the diameter of a proton; however, the gravitational signal he anticipated was not likely to get much greater than the threshold stochastic noise caused by thermal agitation.

It took several years for Weber and his team to begin detecting what they claimed were gravitational wave signals. In 1969 he published results announcing the detection of waves [23]. A year later, Weber claimed that he had discovered many signals that seemed to emanate from the center of our galaxy [24]. This meant that in the center of the Milky Way a lot of stellar mass became energy ($E = mc^2$) in the form of gravitational waves, thus reducing the mass of our galaxy. This "fact" presented the problem that a mass conversion into energy as large as Weber's results implied involved a rapid decrease of the mass that gravitationally keeps our galaxy together. If that were the case, our galaxy would have already been dispersed long ago. Theoretical physicists Sciama, Field, and Rees calculated that the maximum conversion of mass into energy for the galaxy, so as not to expand more than what measurements allowed, corresponded to an upper limit of 200 solar masses per year [25]. However, Weber's measurements implied that a conversion of 1000 solar masses per year was taking place. Something did not fit. Discussions took place to determine what mechanisms could make Weber's measurements possible. Among others, Charles Misner, also from the University of Maryland, put forward the idea that signals, if stemming from the center of the Milky Way, could have originated by gravitational synchotron radiation in narrow angles, so as to avoid the above constraints considered for isotropic emission. Some others, like Peter Kafka of the Max Planck Institute in Munich, claimed in an essay for the Gravity Research Foundation's contest in 1972 (in which he won the second prize) that Weber's measurements, if they were isotropically emitted, and taking into account the inefficiency of bars, would imply a conversion of three million solar masses per year in the center of the Milky Way [26]. It soon became clear that Weber's alleged discoveries were not credible. Weber's frequent observations of gravitational waves related to very sporadic events and raised many suspicions among some scientists. It seemed that Weber was like those who have a hammer in hand and to them everything looks like a nail to hit.

Despite Weber's doubtful measurements, he began to acquire notoriety. In 1971, the famous magazine *Scientific American* invited him to write an article for their readers entitled "The detection of gravitational waves" [27].

Whether it was the amazing—for some—findings of Weber, or doubtful findings of others, or the remarks made by Sciama, Field, Rees, and Kafka, the fact is that many groups of scientists thought it was a good idea to build their own gravitational wave detectors to repeat and improve on Weber's measurements. These first-generation antennas were aluminum cylinders weighing about 1.5 tons and operating at room temperature [28]. Joseph Weber is considered a pioneer in experimental gravitation and therefore he is honored by the American Astronomical Society, which awards every year the Joseph Weber Award for Astronomical Instrumentation.

By the mid-seventies, several detectors were already operative and offered many improvements over Weber's original design; some cylinders were even cooled to reduce thermal noise. These experiments were operating in several places: at Bell Labs Rochester-Holmdel; at the University of Glasgow, Scotland; in an Italo-German joint program in Munich and Frascati; in Moscow; in Tokyo; and at the IBM labs in Yorktown Heights [28]. As soon as these new instruments were put into operation, a common pattern emerged: there were no signals. In the late seventies, everyone except Weber himself agreed that his proclaimed detections were spurious. However, the invalidation of Weber's results urged other researchers to redouble the search for gravitational waves or devise indirect methods of detection.

At that time great pessimism and disappointment reigned among the "seekers" of gravitational waves. However, in 1974 an event occurred that raised hopes. In that year Joseph Hooten Taylor and Alan Russell Hulse found an object in the sky (a binary pulsar) that revealed that an accelerated mass radiated gravitational energy. While this observation did not directly detect gravitational waves, it pointed to their existence. The announcement of the detection of gravitational radiation effects was made in 1979 [29].

This announcement sparked renewed interest in the future discovery of gravitational waves, and urged other researchers to redouble the search for the lost waves and devise other methods of detection. Some were already trying the interferometric method (see Figure 5).

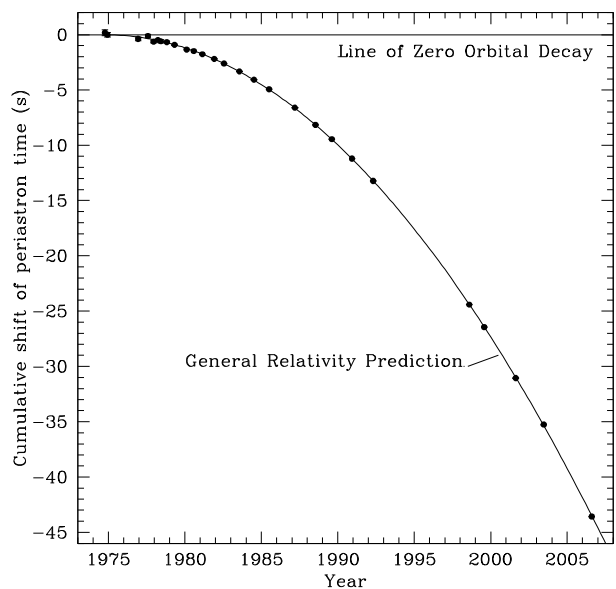

**Figure 5.** Binary Pulsar Advance of the Periastron (point of closest approach of the stars) versus Time. General Relativity predicts this change because of the energy radiated away by gravitational waves. Hulse and Taylor were awarded the Nobel Prize for this observation in 1993. Figure taken from (Living Rev.Rel.11:8, 2008).

To proceed with the description of the method that uses interferometers, it is necessary to know the magnitude of the expected effects of a gravitational wave on matter. This magnitude is properly quantified by the "h" parameter.

**8. The Dimensionless Amplitude, h**

The problem with gravitational waves, as recognized by Einstein ever since he deduced for the first time their existence, is that their effect on matter is almost negligible. Among other reasons, the value of the gravitational constant is very, very small, which makes a possible experimental observation extremely difficult.

Furthermore, not all waves are equal, as this depends on the phenomenon that generates them; nor is the effect of a wave on matter has the same intensity. To evaluate the intensity of the effect that a particular wave produces on matter, a dimensionless factor, denoted by the letter "h," has been defined. The dimensionless amplitude "h" describes the maximum displacement per unit length that would produce waves on an object. To illustrate this definition we refer to Figure 6. This figure shows two particles represented by gray circles. The pair is shown originally spaced by a distance "l" and locally at rest.

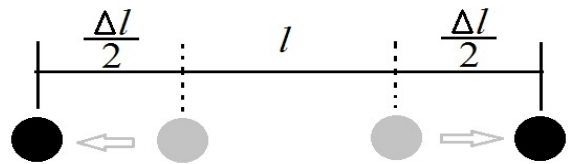

**Figure 6.** Definition of dimensionless amplitude h = Δl/L.

By impinging a gravitational wave perpendicularly on the sheet of paper, both particles are shifted respective to the positions marked by black circles. This shift is denoted by Δl/ 2, which means a relative shift between the pair of particles is now equal to Δl/L ≈ h, where Δl is the change in the spacing between particles due to gravitational wave, l is the initial distance between particles, and h is the dimensionless amplitude. In reality the factor h is more complex and depends upon the geometry of the measurement device, the arrival direction, and the frequency and polarization of the gravitational wave [14]. Nature sets a natural amplitude of h~$10^{-21}$.

This factor h is important when considering the design of a realistic gravitational wave detector. We must mention that the value of h depends on the kind of wave to be detected and this in turn depends on how the wave was produced and how far its source is from an observer. Later we shall return to the subject and the reader shall see the practicality of the factor h.

To identify the sources that produce gravitational waves it is important to consider their temporal behavior. Gravitational waves are classified into three types: stochastic, periodic, and impulsive (bursts) [28]. Stochastic waves contribute to the gravitational background noise and possibly have their origin in the Big Bang. There are also expected stochastic backgrounds due to Black Hole-Black Hole coalescences. These types of waves fluctuate randomly and would be difficult to identify and separate due to the background noise caused by the instruments themselves. However, their identification could be achieved by correlating data from different detectors; this technique applies to other wave types too [14]. The second type of wave, periodic, corresponds to those whose frequency is more or less constant for long periods of time. Their frequency can vary up to a limit (quasiperiodic). For example, these waves may have their origin in binary neutron stars rotating around their center of mass, or from a neutron star that is close to absorb material from another star (accreting neutron star). The intensity of the generated waves depends on the distance from the binary source to the observer. The third type of wave comes from impulsive sources such as bursts that emit pulses of intense gravitational radiation. These may be produced during the creation of Black Holes in a supernova explosion or through the merging of two black holes. The greater the mass, the more intense the signal. They radiate at a frequency inversely proportional to their mass. Such sources are more intense and are expected to have higher amplitudes. Figure 7 shows various examples of possible sources of gravitational waves in which the three different wave types appear in different parts of the spectrum.

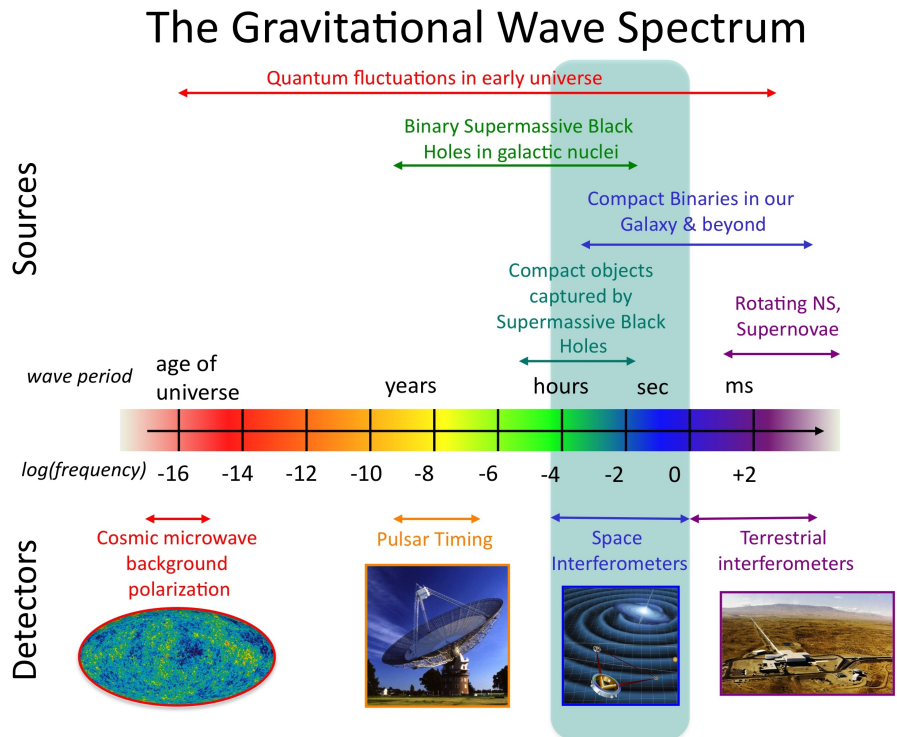

**Figure 7.** Gravitational wave spectrum showing wavelength and frequency along with some anticipated sources and the kind of detectors one might use. Figure credit: NASA Goddard Space Flight Center.

Different gravitational phenomena give rise to different gravitational wave emissions. We expect primordial gravitational waves stemming from the inflationary era of the very early universe. Primordial quantum fields fluctuate and yield space–time ripples at a wide range of frequencies. These could in principle be detected as B-mode polarization patterns in the Cosmic Microwave Background radiation, at large angles in the sky. Unsuccessful efforts have been reported in recent years, due to the difficulty of disentangling the *noisy* dust emission contribution of our own galaxy, the BICEP2 and PLANCK projects. On the other hand, waves of higher frequencies but still very long wavelengths arising from the slow inspiral of massive black holes in the centers of merged galaxies will cause a modified pulse arrival timing, if very stable pulsars are monitored. Pulsar timing also places the best limits on potential gravitational radiation from cosmic string residuals from early universe phase transitions. Other facilities are planned as space interferometers, such as the Laser Interferometer Space Antenna (LISA), which is planned to measure frequencies between 0.03 mHz and 0.1 Hz. LISA plans to detect gravitational waves by measuring separation changes between fiducial masses in three spacecrafts that are supposed to be 5 million kilometers apart! The expected sources are merging of very massive Black Holes at high redshifts, which corresponds to waves emitted when the universe was 20 times smaller than it is today. It should also detect waves from tens of stellar-mass compact objects spiraling into central massive Black Holes that were emitted when the universe was one half of its present size. Last but not least, Figure 7 shows terrestrial interferometers that are planned to detect waves in the frequency from Hertz to 10,000 Hertz. The most prominent facilities are those of LIGO in the USA, VIRGO in Italy, GEO600 in Germany, and KAGRA in Japan, which are all running or expected to run soon. They are just beginning to detect Black Hole mergings, as was the case of the 14 September event [2]. We will go into detail about the interferometric technique later.

## 9. The Origin of the Interferometric Method

It is not known for sure who invented the interferometer method to detect gravitational waves, possibly because the method had several precursors. After all, an idea can arise at the same time among various individuals and this indeed seems to be the case here. However, before going into detail about the historical origins of the method, we shall briefly discuss the basics of this technique.

Figure 8 shows a very simplified interferometer. It consists of a light source (a laser), a pair of reflective mirrors attached to a pair of test masses (not shown in the figure), a beam splitter (which can be a semi-reflecting mirror or half-silvered mirror), and a light detector or photodetector.

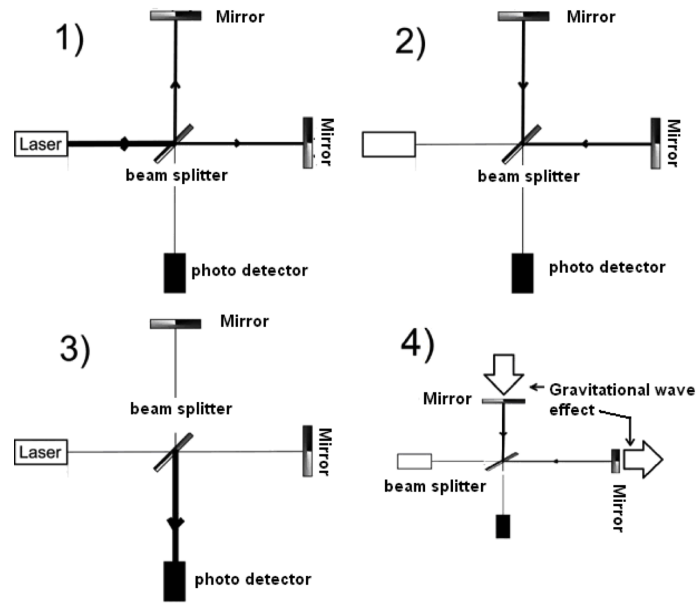

**Figure 8.** Schematic of an interferometer for detecting gravitational waves.

The laser source emits a beam of monochromatic light (i.e., at a single frequency) that hits the beam splitter surface. This surface is partially reflective, so part of the light is transmitted through to the mirror at the right side of the diagram while some is reflected to the mirror at the upper side of the sketch (Figure 8.1). Then, as seen in Figure 8.2, both beams recombine when they meet at the splitter and the resulting beam is reflected toward the detector (Figure 8.3). Finally, the photodetector measures the light intensity of the recombined beam. This intensity is proportional to the square of the height of the recombined wave.

Initially, both reflecting mirrors are positioned at nearly the same distance from the beam splitter. In reality what is needed is that the interferometer is locked on a dark fringe. Deviations from a dark fringe are then measured with the passage of a gravitational wave.

If the distance between one of the mirrors to the light splitter varies by an amount Δl with respect to distance to the same splitter of the second mirror, then the recombined beam will change its intensity. From measuring the intensity change of the recombined light beam, it is possible to obtain Δl.

When a gravitational wave passes through the interferometer at a certain direction, for example perpendicular to the plane where the pair of mirrors lies, both mirrors shift positions. One of the mirrors slightly reduces its distance to the beam splitter, while the second mirror slightly increases its distance to the splitter (see Figure 8.4). The sum of the two displacements is equal to Δl. The photodetector records a variation in the intensity of the recombined light, thereby detecting the effect of gravitational waves.

A very important specific feature of the interferometer effectiveness is given by the length of its arms. This is the distance "l" between the wave splitter and its mirrors. On the other hand, the wavelength of the gravitational wave sets the size of the detector L needed. The optimal size of the arms turns out to be one-fourth of the wavelength. For a typical gravitational wave frequency of 100 Hz, this implies L = 750 km, which is actually too long to make except by folding the beams back and forth via the Fabry–Pérot technique, which helps to achieve the desired optimal size. In practice it is of the utmost importance to have an interferometer with very long arms matching the frequencies one plans to observe.

The importance of a long arm "l" is easy to explain if we remember the definition of the dimensionless amplitude h = Δl/L, which we already discussed above (see Figure 6). If a gravitational wave produces a displacement Δl for the distance between the mirrors, according to the definition of h, the resulting change Δl will be greater the longer the interferometer arm l is, since they are directly proportional (Δl = L × h). Therefore, the reader can notice that interferometric arm lengths must be tailored depending on what type of gravitational source is intended for observation. The explanation just given corresponds to a very basic interferometer, but in actuality these instruments are more complex. We will come back to this.

**10. Genesis of the Interferometer Method (Or, Who Deserves the Credit?)**

Let us now turn our attention to the origin of the interferometer method. The first explicit suggestion of a laser interferometer detector was outlined in the former USSR by Gertsenshtein and Pustovoid in 1962 [30]. The idea was not carried out and eventually was resurrected in 1966 behind the "Iron Curtain" by Vladimir B. Braginskiĭ, but then again fell into oblivion [31].

Some year before, Joseph Weber returned to his laboratory at the University of Maryland after having attended the 1957 meeting at Chapel Hill. Weber came back bringing loads of ideas. Back at his university, Weber outlined several schemes on how to detect gravitational waves. As mentioned, one of those was the use of a resonant "antenna" (or cylinder), which, in the end, he finally built. However, among other various projects, he conceived the use of interferometer detectors. He did not pursue this conception, though, and the notion was only documented in the pages of his laboratory notebook [28] (p. 414). One can only say that history produces ironies.

By the end of 1959 Weber began the assembly of his first "antenna" with the help of his students Robert L. Forward and David M. Zipoy [32]. Forward would later (in 1978) turn out to be the first scientist to build an interferometric detector [33].

In the early seventies Robert L. Forward, a former student of Joseph Weber at that time working for Hughes Research Laboratory in Malibu, California, decided, with the encouragement of Rainer Weiss, to build a laboratory interferometer with Hughes' funds. Forward's interest in interferometer detectors had evolved some years before when he worked for Joseph Weber at his laboratory in the University of Maryland on the development and construction of Weber's antennas.

By 1971 Forward reported the design of the first interferometer prototype (which he called a "Transducer Laser"). In his publication Forward explained, "The idea of detecting gravitational radiation by using a laser to measure the differential motion of two isolated masses has often been suggested in past[5]" The footnote reads, "To our knowledge, the first suggestion [of the interferometer device] was made by J. Weber in a telephone conversation with one of us (RLF) [Forward] on 14 September 1964" [34].

After 150 hours of observation with his 8.5-m arms interferometer, Robert Forward reported "an absence of significant correlation between the interferometer and several Weber bars detectors, operating at Maryland, Argonne, Glasgow and Frascati." In short, Robert Forward did not observe gravitational waves. Interestingly, in the acknowledgments of his article, Forward recognizes the advice of Philip Chapman and Weiss [35].

Also in the 1970s, Weiss independently conceived the idea of building a Laser Interferometer, inspired by an article written by Felix Pirani, the theoretical physicist who, as we have already mentioned, developed in 1956 the necessary theory to grow the conceptual framework of the method [12]. In this case it was Weiss who developed the method. However, it was not only Pirani's paper that influenced Weiss; he also held talks with Phillip Chapman, who had glimpsed, independently, the same scheme [36]. Chapman had been a member of staff at MIT, where he worked on electro-optical systems and gravitational theory. He left MIT to join NASA, where he served from 1967 to 1972 as a scientist–astronaut (he never went to space). After leaving NASA, Chapman was employed as a researcher in laser propulsion systems at Avco Everett Research Laboratory in Malibu, California. It was at this time that he exchanged views with Weiss. Chapman subsequently lost interest in topics related to gravitation and devoted himself to other activities.

In addition, Weiss also held discussions with a group of his students during a seminar on General Relativity he was running at MIT. Weiss gives credit to all these sources in one of his first publications on the topic: "The notion is not new; It has appeared as a gedanken experiment in F.A.E. Pirani's studies of the measurable properties of the Riemann tensor. However, the realization that with the advent of lasers it is feasible to detect gravitational waves by using this technique [interferometry], grew out of an undergraduate seminar that I ran at MIT several years ago, and has been independently discovered by Dr. Phillip Chapman of the the National Aeronautics and Space Administration, Houston" [36].

Weiss recalled, in a recent interview, that the idea was incubated in 1967 when he was asked by the head of the teaching program in physics at MIT to give a course of General Relativity. At that time Weiss's students were very interested in knowing about the "discoveries" made by Weber in the late sixties. However, Weiss recalls that "I couldn't for the life of me understand the thing he was doing" and "I couldn't explain it to the students." He confesses "that was my quandary at the time" [37].

A year later (in 1968) Weiss began to suspect the validity of Weber's observations because other groups could not verify them. He thought something was wrong. In view of this he decided to spend a summer in a small cubicle and worked the whole season on one idea that had occurred to him during discussions with his students at the seminar he ran at MIT [37].

After a while, Weiss started building a 1.5-m long interferometer prototype, in the RLE (Research Laboratory of Electronics) at MIT using military funds. Some time later, a law was enacted in the United States (the "Mansfield amendment"), which prohibited Armed Forces financing projects that were not of strictly military utility. Funding was suddenly suspended. This forced Weiss to seek financing from other U.S. government and private agencies [37].

**11. Wave Hunters on a Merry-Go-Round (GEO)**

In 1974, NSF asked Peter Kafka of the Max Planck Institute in Munich to review a project. The project was submitted by Weiss, who requested $53,000 in funds for enlarging the construction of a prototype interferometer with arms nine meters in length [38]. Kafka agreed to review the proposal. Being a theoretician himself, Kafka showed the proposal documents to some experimental physicists at his institute for advice. To Kafka's embarrassment, the local group currently working on Weber bars became very enthusiastic about Weiss's project and decided to build their own prototype [39], headed by Heinz Billing.

This German group had already worked in collaboration with an Italian group in the construction of "Weber antennas." The Italian–German collaboration found that Weber was wrong [37]. The Weiss proposal fell handily to the Germans as they were in the process of designing a novel Weber antenna that was to be cooled to temperatures near absolute zero to reduce thermal noise in the new system. However, learning of Weiss's proposal caused a shift in the research plans of the Garching group. They made the decision to try the interferometer idea. Germans contacted Weiss for advice and they also offered a job to one of his students on the condition that he be trained on the Weiss 1.5-m prototype. Eventually Weiss sent David Shoemaker, who had worked on the MIT prototype, to join the Garching group. Shoemaker later helped to build a German 3-m prototype and later a 30-m interferometer [40]. This interferometer in Garching served for the development of noise suppression methods that would later be used by the LIGO project.

It is interesting to mention that Weiss's proposal may seem modest nowadays (9-m arms), but he already had in mind large-scale interferometers. His prototype was meant to lead to a next stage featuring a one-kilometer arm length device, which his document claimed it would be capable of detecting waves from the Crab pulsar (PSR B0531 + 21) if its periodic signal were integrated over a period of months. Furthermore, the project envisaged a future development stage where long baseline interferometers in outer space could eventually integrate Crab pulsar gravitational waves in a matter of few hours .

NSF then supported Weiss's project and funds were granted in May 1975 [38,39].

At that time Ronald Drever, then at the University of Glasgow, attended the International School of Cosmology and Gravitation in Erice, *La Città della Scienza*, Sicily in March 1975. There, a lecture entitled "Optimal detection of signals through linear devices with thermal source noises, an application to the Munich–Frascati Weber-type gravitational wave detectors" was delivered by the same Peter Kafka of Munich [41]. His lecture was again very critical of Weber's results and went on to showing that the current state of the art of Weber bars including the Munich–Frascati experiment, was far from the optimal sensitivity required for detection. In fact, the conclusion of his notes reads: "It seems obvious that only a combination of extremely high quality and extremely low temperature will bring resonance detectors [Weber bars] near the range where astronomical work is possible. Another way which seems worth exploring is *Laser interferometry with long free mass antennas*" (emphasis ours) [41].

Ronald Drever was part of Kafka's lecture audience. The lecture probably impressed him as he started developing interferometric techniques on his return to Glasgow. He began with simple tasks, a result of not having enough money. One of them was measuring the separation between two massive bars with an interferometer monitoring the vibrations of aluminum bar detectors. The bars were given to him by the group at the University of Reading, United Kingdom. The bars were two halves of the Reading group's split bar antenna experiment [42]. By the end of the 1970s he was leading a team at Glasgow that had completed a 10-meter interferometer. Then in 1979 Drever was invited to head up the team at Caltech, where he accepted a part-time post. James Hough took his place in Glasgow.

Likewise, in 1975 the German group at Munich (Winkler, Rüdiger, Schilling, Schnupp, and Maischberger), under the leadership of Heinz Billing, built a prototype with an arm length of 3 m [43]. This first prototype displayed unwanted effects such as laser frequency instabilities, lack of power, a shaky suspension system, etc. The group worked hard to reduce all these unwanted effects by developing innovative technologies that modern-day gravitational interferometers embraced. In 1983 the same group, now at the Max Planck Institute of Quantum Optics (MPQ) in Garching, improved their first prototype by building a 30 m arm length instrument [44]. To "virtually" increase the optical arm length of their apparatus, they "folded" the laser beam path by reflecting the beam backwards and forwards between the mirrors many times, a procedure that is known as "delay line." In Weiss's words, "the Max Planck group actually did most of the very early interesting development. They came up with a lot of what I would call the practical ideas to make this thing [gravitational interferometers] better and better" [45].

After a couple of years of operating the 30-m model, the Garching group was prepared to go for Big-Science. In effect, in June 1985, they presented a document "Plans for a large Gravitational wave antenna in Germany" at the Marcel Grossmann Meeting in Rome [46]. This document contains the first detailed proposal for a full-sized interferometer (3 km). The project was submitted for funding to the German authorities but there was not sufficient interest in Germany at that time, so it was not approved.

In the meantime, similar research was undertaken by the group at Glasgow, now under Jim Hough after Drever's exodus to Caltech. Following the construction of their 10-m interferometer, the Scots decided in 1986 to take a further step by designing a Long Baseline Gravitational Wave Observatory [47]. Funds were asked for, but their call fell on deaf ears.

Nevertheless, similar fates bring people together, but it is still up to them to make it happen. So, three years later, the Glasgow and the Garching groups decided to unite efforts to collaborate in a plan to build a large detector. It did not take long for both groups to jointly submit a plan for an underground 3-km installation to be constructed in the Harz Mountains in Germany, but again their proposal was not funded [48]. Although reviewed positively, a shortage of funds on both ends (the British Science and Engineering Research Council (SERC) and the Federal Ministry of Research and Technology (Bundesministerium für Forschung und Technologie BMFT)) prevented the approval. The reason for the lack of funds for science in Germany was a consequence of the German re-unification (1989–1990), as there was a need to boost the Eastern German economy; since private, Western funds lagged, public funds were funneled to the former German Democratic Republic. Incidentally the Harz Mountains are the land of German fairy tales.

In spite of this disheartening ruling, the new partners decided to try for a shorter detector and compensate by employing more advanced and clever techniques [49]. A step forward was finally taken in 1994 when the University of Hanover and the State of Lower Saxony donated ground to build a 600-m instrument in Ruthe, 20 km south of Hanover. Funding was provided by several German and British agencies. The construction of GEO 600 started on 4 September 1995.

The following years of continuous hard work by the British and Germans brought results. Since 2002, the detector has been operated by the Center for Gravitational Physics, of which the Max Planck Institute is a member, together with Leibniz Universität in Hanover and Glasgow and Cardiff Universities.

The first stable operation of the Power Recycled interferometer was achieved in December 2001, immediately followed by a short coincidence test run with the LIGO detectors, testing the stability of the system and getting acquainted with data storage and exchange procedures. The first scientific data run, again together with the LIGO detectors, was performed in August and September of 2002. In November 2005, it was announced that the LIGO and GEO instruments began an extended joint science run [50]. In addition to being an excellent observatory, the GEO 600 facility has served as a development and test laboratory for technologies that have been incorporated in other detectors all over the world.

**12. The (Nearly)…Very Improbable Radio Gravitational Observatory—VIRGO**

In the late 70s, when Allain Brillet was attracted to the detection of gravitational waves, the field was ignored by a good number of his colleagues after the incorrect claims of Joe Weber. However, Weiss's pioneering work on laser interferometers in the early 1970s seemed to offer more chances of detection beyond those of Weber bars.

Brillet's interest in the field started during a postdoc stay at the University of Colorado, Boulder under Peter L. Bender of the Laboratory of Astrophysics, who, together with Jim Faller, first proposed the basic concept behind LISA (the Laser Interferometer Space Antenna). Brillet also visited Weiss at MIT in 1980 and 1981 where he established good links with him that produced, as we shall see, a fruitful collaboration in the years to come.

Upon Brillet's return to France in 1982 he joined a group at Orsay UPMC that shared the same interests. This small group (Allain Brillet, Jean Yves Vinet, Nary Man, and two engineers) experienced difficulties and had to find refuge in the nuclear physics department of the Laboratoire de Physique de l'Institute Henri Poincaré, led by Philippe Tourrenc [51].

On 14 November 1983 a meeting on Relativity and Gravitation was organized by the Direction des Etudes Recherches et Techniques de la Délégation Générale pour l'Armement. One of the objectives of the meeting was the development of a French project of gravitational wave detectors [52]. There, Brillet gave a lecture that advocated for the use of interferometers as the best possible detection method [53]. His lecture raised some interest, but French agencies and academic departments were not willing to invest money or personnel in this area. The technology was not yet available, mainly in terms of power laser stability, high-quality optical components, and seismic and thermal noise isolation. In addition, at that point in time there was no significant experimental research on gravitation in France.

However, interest in gravitational wave detection began to change when Hulse and Taylor demonstrated the existence of gravitational waves. It was at the Marcel Grossman meeting of 1985 in Rome that Brillet met Adalberto Giazotto, an Italian scientist working at the Universita di Pisa on the development of suspension systems. At that meeting Giazotto put forward his ideas and the first results of his super-attenuators, devices that serve as seismic isolators to which interferometer mirrors could be attached. During the same meeting Jean-Yves Vinet (Brillet's colleague) gave a talk about his theory of recycling, a technique invented by Ronald Drever to reduce by a large factor the laser power required by gravitational interferometers. Conditions for a partnership were given.

Both scientists then approached the research leaders of a German project (the Max Planck Institute of Quantum Optics in Garching), hoping to collaborate on a big European detector, but they were told that their project was "close to being financed" and the team at Garching did not accept the idea of establishing this international collaboration, because it "would delay project approval" [51].

So they decided to start their own parallel project, the VIRGO Interferometer, named for the cluster of about 1500 galaxies in the Virgo constellation about 50 million light-years from Earth. As no terrestrial source of gravitational wave is powerful enough to produce a detectable signal, VIRGO must observe far enough out into the universe to see many of the potential source sites; the Virgo Cluster is the nearest large cluster.

At that time Brillet was told that CNRS (Le Centre National de la Recherche Scientifique) would not be able to finance VIRGO's construction on the grounds that priority was given to the Very Large Telescope in Chile. Even so, both groups (Orsay and Pisa) did not give in to dismay and continued their collaboration; in 1989 they were joined by the groups of Frascati and Naples. This time they decided to submit the VIRGO project to the CNRS (France) and the INFN (Istituto Nazionale di Fisica Nucleare, Italy) [51].

The VIRGO project was approved in 1993 by the French CNRS and in 1994 by the Italian INFN. The place chosen for VIRGO was the alluvial plain of Cascina near Pisa. The first problem INFN encountered was persuading the nearly 50 land title holders to cooperate and sell their parcels to the government. Gathering the titles took a long time. The construction of the premises started in 1996. To complicate matters further, VIRGO's main building was constructed on a very flat alluvial plain, so it was vulnerable to flooding. That took additional time to remedy [54].

From the beginning it was decided to use the VIRGO interferometer as its own prototype, in contrast to LIGO, which used MIT and Caltech and the German–British (Geo 600) installations to test previous designs before integrating them into the main instrument. This strategy was decided on the grounds that it would be faster to solve problems in actual size directly, rather than spend years on a smaller prototype and only then face the real difficulties.

Between 1996 and 1999, VIRGO had management problems as the construction was handled by an association of separate laboratories without a unified leadership, so it was difficult to ensure proper coordination [43]. As a result, in December 2000 the French CNRS and the Italian INFN created the European Gravitational Observatory (EGO consortium), responsible for the VIRGO site, the construction, maintenance, and operation of the detector, and its upgrades.

The construction of the initial VIRGO detector was completed in June 2003 [55]. It was not until 2007 that VIRGO and LIGO agreed to join in a collaborative search for gravitational waves. This formal agreement between VIRGO and LIGO comprises the exchange of data, and joint analysis and co-authorship of all publications concerned. Several joint data-taking periods followed between 2007 and 2011.

Even though a formal cooperation has been established, continued informal cooperation has been running for years ever since Alain Brillet visited the MIT laboratories back in 1980–1981. As a matter of fact, VIRGO and LIGO have exchanged a good number of students and postdocs. Just to name one, David Shoemaker, the current MIT LIGO Laboratory Director, received his PhD on the

Nd-YAG lasers and recycling at Orsay before joining LIGO. Also, in 1990 Jean Yves Vinet provided LIGO with a computer simulation program necessary to specify its optical system. LIGO adapted this computer code.

The year 2016 will be an important milestone for the construction of the advanced VIRGO detector. After a months-long commissioning period, the advanced VIRGO detector will join the two advanced LIGO detectors ("aLIGO") for a first common data-taking period that should include on the order of one gravitational wave event per month. With all three detectors operating, data can be further correlated and the direction of the gravity waves' source should be much more localized.

### 13. The Origin of the LIGO Project

In the summer of 1975 Weiss went to Dulles Airport in Washington, D.C. to pick up Kip Thorne, a renowned theoretical physicist from Caltech. The reason for visiting Washington was attending a NASA meeting on uses of space research in the field of cosmology and relativity. (One of us, GFS, attended this meeting as a young postdoc and remembers a presentation by a tired Rai Weiss on the concept of a "Laser Interferometer Space Array" for detecting gravitational waves. It was a very naïve and ambitious space project presentation and only in about 2035 (60 years later) does it appear likely to be a working realization. The meeting did open up to me the idea of doing science in space.) Weiss recalls, "I picked Kip up at the airport on a hot summer night when Washington, D.C., was filled with tourists. He did not have a hotel reservation so we shared a room for the night" [37].

They did not sleep that night because both spent the night discussing many topics, among them how to search for gravitational waves.

Weiss remembers that night "We made a huge map on a piece of paper of all the different areas in gravity. Where was there a future? Or what was the future, or the thing to do?" [56]. Thorne decided that night that the thing they ought to do at Caltech was interferometric gravitational wave detection. However, he would need help from an experimental physicist.

Thorne first thought of bringing to the United States his friend Vladimir B. Braginskiĭ, a Russian scientist who had closely worked with him and, moreover, had already acquired experience in the search for gravitational waves [57]. However, the Cold War prevented his transfer. Meanwhile, Weiss suggested another name, Ronald Drever. Weiss had only known Drever from his papers, not in person. Drever was famous for the Hughes–Drever Experiments, spectroscopic tests of the isotropy of mass and space confirming the Lorentz invariance aspects of the theory of Relativity, and had also been the leader of the group that built a "Weber cylinder" at the University of Glasgow [58]. At that time Drever was planning the construction of an interferometer.

In 1978 Thorne offered a job to Drever at Caltech for the construction of an interferometer. Drever accepted the offer in 1979, dividing his time between the Scottish university and Caltech. Hiring Drever half-time soon paid dividends because in 1983 he had already built his first instrument at Caltech, an interferometer whose "arms" measured 40 m. The instrument was noisier than expected and new ingenious solutions ranging from improving seismic isolation and laser power increase to stabilization were attempted. In 1983 Drever began full-time work at Caltech with the idea of gradually improving and increasing the size of the prototype as it was built and run. In contrast to the Caltech apparatus, as already mentioned, at MIT Weiss had built a modest 1.5-m prototype with a much smaller budget than the Californian instrument. In late 1979 the NSF granted modest funds to the Caltech interferometer group and gave a much smaller amount of money to the MIT team. Soon Drever and Weiss began to compete to build more sensitive and sophisticated interferometers.

The sensitivity of the interferometers can be enhanced by boosting the power of the lasers and increasing the optical path of the light beam as it travels through the interferometer arms.

To increase the sensitivity of the interferometer, Weiss put forward the use of an optical delay line. In the optical delay method, the laser light passes through a small hole in an adjacent wave divider mirror and the beam is reflected several times before emerging through the inlet port.

Figure 9 shows a simplified diagram of the method. In this figure only a couple of light "bounces" are shown between the mirrors to maintain clarity of the scheme, but in reality the beam is reflected multiple times. This method effectively increases the length of the interferometer.

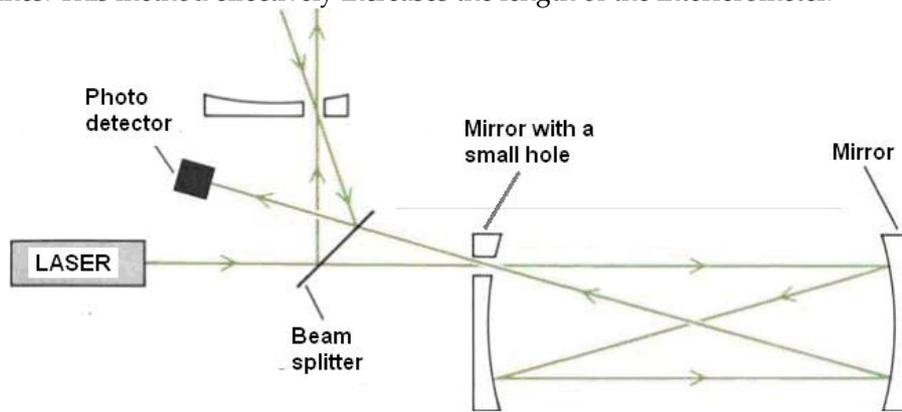

**Figure 9.** Optical delay method.

In the meantime Drever developed an arrangement that utilized "Fabry–Pérot cavities." In this method the light passes through a partially transmitting mirror to enter a resonant cavity flanked at the opposite end by a fully reflecting mirror. Subsequently, the light escapes through the first mirror, as shown in Figure 10.

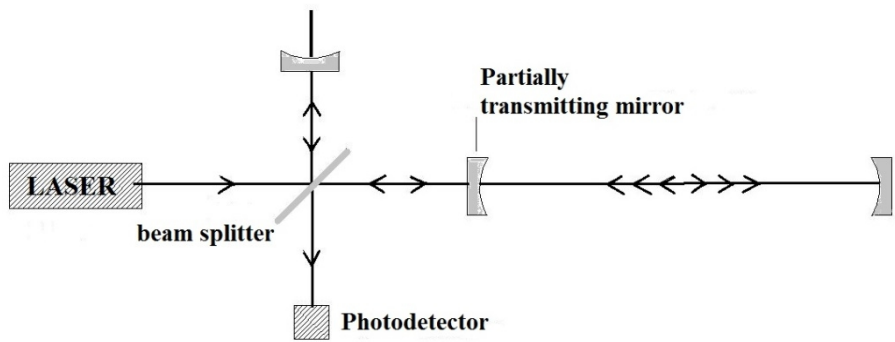

**Figure 10.** Fabry–Pérot method.

As already mentioned, Weiss experimented with an interferometer whose two L-shaped "arms" were 1.5 m long. Drever, meanwhile, had already built and operated a 40-m interferometer. With the Caltech group appearing to be taking the lead, Weiss decided in 1979 to "do something dramatic." That year Weiss held talks with Richard Isaacson, who at the time served as program director of the NSF Gravitational Physics division. Isaacson had a very strong professional interest in the search for gravitational waves as he had developed a mathematical formalism to approximate gravitational waves solutions from Einstein's equations of General Relativity in situations where the gravitational fields are very strong [59]. Weiss offered to conduct a study in collaboration with industry partners to determine the feasibility and cost of an interferometer whose arms should measure in kilometers. In turn Isaacson receive a document to substantiate a device on a scale of kilometers, with possible increased funding from the NSF. The study would be funded by NSF. The study by Weiss and colleagues took three years to complete. The produced document was entitled "A study of a long Baseline Gravitational Wave Antenna System," co-authored by Peter Saulson and Paul Linsay [60]. This fundamental document is nowadays known as the "The Blue Book" and covers very many important issues in the construction and operation of such a large interferometer. The Blue Book was submitted to the NSF in October 1983. The proposed budget was just under $100 million to build two instruments located in the United States.

Before "The Blue Book" was submitted for consideration to the NSF, Weiss met with Thorne and Drever at a Relativity congress in Italy. There they discussed how they could work together—this was mandatory, because the NSF would not fund two megaprojects on the same subject and with the same objective. However, from the very beginning it was clear that Drever did not want to collaborate with Weiss, and Thorne had to act as mediator. In fact, the NSF settled matters by integrating the MIT and Caltech groups together in a "shotgun wedding" so the "Caltech–MIT" project could be jointly submitted to the NSF [56].

**14. The LIGO Project**

The Caltech–MIT project was funded by NSF and named the "Laser Interferometer Gravitational-Wave Observatory," known by its acronym LIGO. The project would be led by a triumvirate of Thorne, Weiss, and Drever. Soon interactions between Drever and Weiss became difficult because, besides the strenuous nature of their interaction, both had differing opinions on technical issues.

During the years 1984 and 1985 the LIGO project suffered many delays due to multiple discussions between Drever and Weiss, mediated when possible by Thorne. In 1986 the NSF called for the dissolution of the triumvirate of Thorne, Drever, and Weiss. Instead Rochus E. Vogt was appointed as a single project manager [61].

In 1988 the project was finally funded by the NSF. From that date until the early 1990s, project progress was slow and underwent a restructuring in 1992. As a result, Drever stopped belonging to the project and in 1994 Vogt was replaced by a new director, Barry Clark Barish, an experimental physicist who was an expert in high-energy physics. Barish had experience in managing big projects in physics. His first activity was to review and substantially amend the original five-year old NSF proposal. With its new administrative leadership, the project received good financial support. Barish's plan was to build the LIGO as an evolutionary laboratory where the first stage, "initial LIGO" (or iLIGO), would aim to test the concept and offer the possibility of detecting gravitational waves. In the second stage ("aLIGO" or advanced LIGO), wave detection would be very likely (see Figure 11).

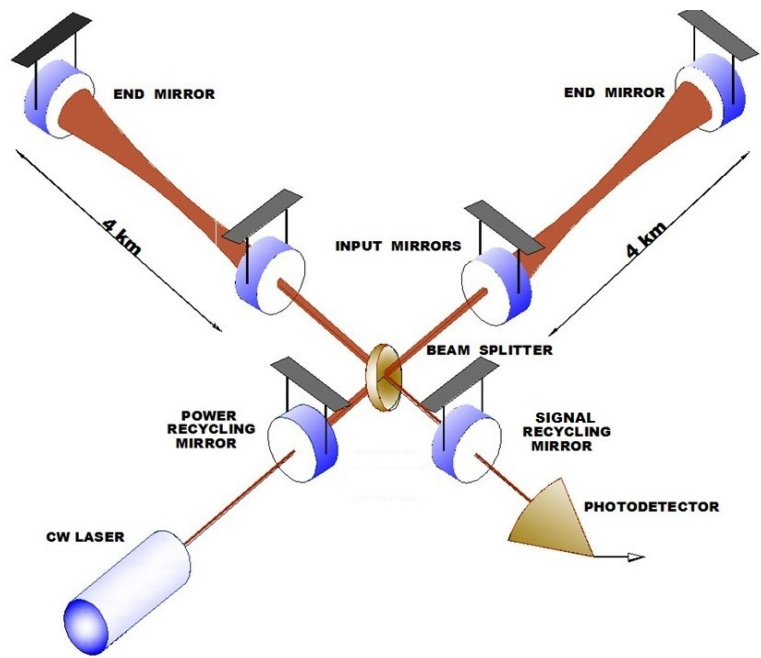

**Figure 11.** Advanced LIGO interferometer design concept. Figure made after T.F. Carruthers and D.H. Reitze, "LIGO," *Optics & Photonics News*, March 2015.

Two observatories, one in Hanford in Washington State and one in Livingston, Louisiana, would be built. Construction began in late 1994 and early 1995, respectively, and ended in 1997. Once the construction of the two observatories was complete, Barish suggested two organizations to be funded: the laboratory LIGO and scientific collaboration LIGO (LIGO Scientific Collaboration (LSC)). The first of these organizations would be responsible for the administration of laboratories. The second organization would be a scientific forum headed by Weiss, responsible for scientific and technological research. LSC would be in charge of establishing alliances and scientific collaborations with Virgo and GEO600.

Barish's idea to make LIGO an evolutionary apparatus proved in the end to pay dividends. The idea was to produce an installation whose parts (vacuum system, optics, suspension systems, etc.) could always be readily improved and buildings that could house those ever-improving interferometer components.

In effect, LIGO was incrementally improved by advances made in its own laboratories and those due to associations with other laboratories (VIRGO and GEO600). To name some: Signal-recycling mirrors were first used in the GEO600 detector, as well as the monolithic fiber-optic suspension system that was introduced into advanced LIGO. In brief, LIGO detection was the result of a worldwide collaboration that helped LIGO evolve into its present remarkably sensitive state.

### 15. Looking Back Over the Trek

The initial LIGO operated between 2002 and 2010 and did not detect gravitational waves. The upgrade of LIGO (advanced LIGO) began in 2010 to replace the detection and noise suppression and improve stability operations at both facilities. This upgrade took five years and had contribution from many sources. For example, the seismic suspension used in aLIGO is essentially the design that has been used in VIRGO since the beginning. While advanced VIRGO is not up and running yet, There have been many technological contributions from both LSC scientists on the European side and VIRGO.

aLIGO began in February 2015 [62]. The team operated in "engineering mode"—that is, in test mode—and in late September began scientific observation [63]. It did not take many days for LIGO to detect gravitational waves [2]. Indeed, LIGO detected the collision of two black holes of about 30 solar masses collapsed to 1300 million light years from Earth.

Even the latest search for gravitational waves was long and storied. The upgrade to aLIGO cost $200 million, and preparing it took longer than expected, so the new and improved instrument's start date was pushed back to 18 September 2015.

### 16. The Event: 14 September 2015

On Sunday September 13th the LIGO team performed a battery of last-minute tests. "We yelled, we vibrated things with shakers, we tapped on things, we introduced magnetic radiation, we did all kinds of things," one of the LIGO members said. "And, of course, everything was taking longer than it was supposed to." At four in the morning, with one test still left to do—a simulation of a truck driver hitting his brakes nearby—we stopped for the night. We went home, leaving the instrument to gather data in peace and quiet. The signal arrived not long after, at 4:50 a.m. local time, passing through the two detectors within seven milliseconds of each other. It was four days before the start of Advanced LIGO's first official run. It was still during the time meant for engineering tests, but nature did not wait.

The signal had been traveling for over a billion years, coming from a pair of 30-solar mass black holes orbiting around their common center of mass and slowly drawing into a tighter and tighter orbit from the energy being lost by gravitational radiation. The event was the end of this process — formed from final inspiral, the merger of the black holes to form a larger one, and the ring down of that new massive black hole. All of this lasted mere thousandths of a second, making the beautiful signals seen by both aLIGO detectors which immediately answered several questions: Are there gravitational waves and can we detect them? Is General Relativity likely right for strong fields? Are large black holes common?

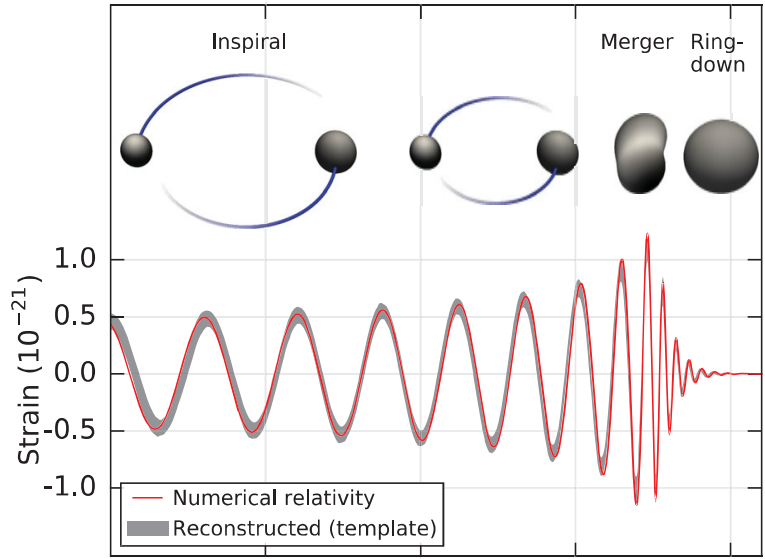

**Figure 12.** Showing the phases: binary orbits inspiraling, black holes merging, and final black hole ringing down to spherical or ellipsoidal shape [2].

The waveform detected by both LIGO observatories matched the predictions of General Relativity for a gravitational wave emanating from the inward spiral and merger of a pair of black holes of around 36 and 29 solar masses and the subsequent "ringdown" of the single resulting black hole. See Figures 12 and 13 for the signals and GR theoretical predictions.

The signal was named GW150914 (from "Gravitational Wave" and the date of observation). It appeared 14 September 2015 and lasted about 0.2 s. The estimated distance to the merged black holes is 410 $^{+160}_{-180}$ Mpc or 1.3 billion light years, which corresponds to a redshift of about z = 0.09. The interesting thing is that the estimated energy output of the event in gravitational waves is about $(3.0+/-0.5)$ $M_\odot \times c^2$. That is three times the rest mass of our sun converted into energy.

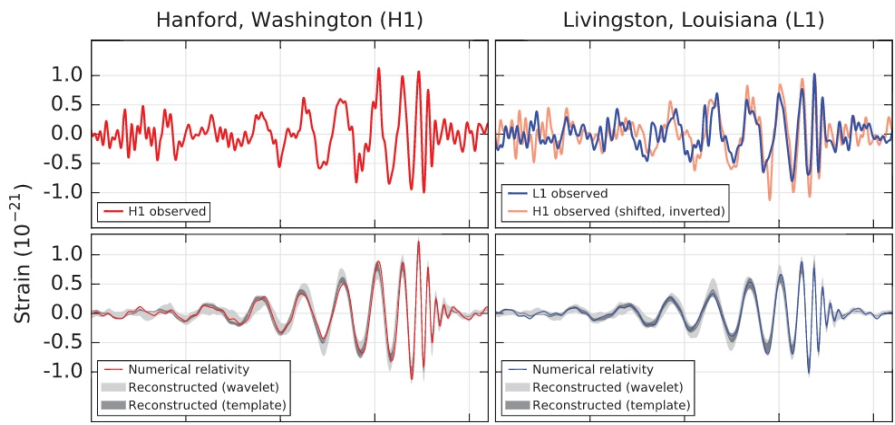

**Figure 13.** LIGO measurement of the gravitational waves at the Livingston (right) and Hanford (left) detectors, compared with the theoretical predicted values from General Relativity [2].

More recently, the aLIGO team announced the detection and analysis of another binary black hole merger event, GW151226. This event is apparently the merger of a 14 solar mass black hole with a 7.5 solar mass black hole, again at a distance of about 440 Mpc (about 1.4 billion light years). An interesting feature is that one of the black holes is measured to have significant spin, s = 0.2. There is also a report of a candidate event that occurred between the two confirmed events and had black holes and energy intermediate to the two confirmed events. These additional reports also show the beginning of measuring rates and the distribution of black hole binary systems. It has long been expected that as aLIGO improves its design sensitivity (about three times better than this first run), that a stable set of events would include the merger of binary neutron stars, for whose inventory we have better estimates. These early events predict that there will be a whole distribution of gravitational wave events observed in the future, from neutron star mergers, black hole mergers, and even neutron star–black hole mergers, see Figure [14]

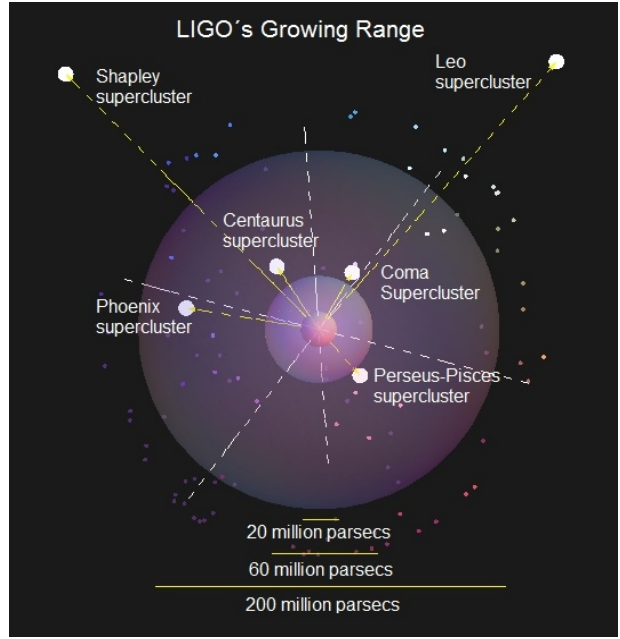

**Figure 14.** Showing the initial range of LIGO and the anticipated range of aLIGO. The volume is much greater and the anticipated rate of events and detections are expected to scale up with the volume. Note the large number of galaxies included in the observational volume. The anticipated factor of 3 in sensitivity should correspondingly increase the event rate by up to nine times.

Albert Einstein originally predicted the existence of gravitational waves in 1916, based upon General Relativity, but wrote that it was unlikely that anyone would ever find a system whose behavior would be measurably influenced by gravitational waves. He was pointing out that the waves from a typical binary star system would carry away so little energy that we would never even notice that the system had changed — and that is true. The reason we can see it from the two black holes is that they are closer together than two stars could ever be. The black holes are so tiny and yet so massive that they can be close enough together to move around each other very, very rapidly. Still, to get such a clear signal required a very large amount of energy and the development of extraordinarily sensitive instruments. This clearly settles the argument about whether gravitational waves really exist; one major early argument was about whether they carried any energy. They do! That was proved strongly and clearly.

Some analyses have been carried out to establish whether or not GW150914 matches with a binary black hole configuration in General Relativity [64]. An initial consistency test encompasses the mass and spin of the end product of the coalescence. In General Relativity, the final black hole product of a binary coalescence is a Kerr black hole, which is completely described by its mass and spin. It has been verified that the remnant mass and spin from the late-stage coalescence deduced by numerical relativity simulations, inferred independently from the early stage, are consistent with each other, with no evidence for disagreement from General Relativity. There is even some data on the ring down phase, but we can hope for a better event to provide quality observations to test this phase of General Relativity.

GW150914 demonstrates the existence of black holes more massive than $\simeq 25 M_\odot$, and establishes that such binary black holes can form in nature and merge within a Hubble time. This is of some surprise to stellar theorists, who predicted smaller mass black holes would be much more common.

## 17. Conclusions

This observation confirms the last remaining unproven prediction of General Relativity (GR)—gravitational waves—and validates its predictions of space–time distortion in the context of large-scale cosmic events (known as strong field tests). It also inaugurates the new era of gravitational-wave astronomy, which promises many more observations of interesting and energetic objects as well as more precise tests of General Relativity and astrophysics. While it is true that we can never rule out deviations from GR at the 100% level, all three detections so far agree with GR to an extremely high level (>96%). This will put constraints on some non-GR theories and their predictions.

With such a spectacular early result, others seem sure to follow. In the four-month run, 47 days' worth of coincident data was useful for scientific analysis, i.e., this is data taken when both LIGOs were in scientific observation mode. The official statement is that these 47 days' worth of data have been fully analyzed and no further signals lie within them. We can expect many more events once the detectors are running again.

For gravitational astronomy, this is just the beginning. Soon, aLIGO will not be alone. By the end of the year VIRGO, a gravitational-wave observatory in Italy, should be operating to join observations and advanced modes. Another detector is under construction in Japan and talks are underway to create a fourth in India. Most ambitiously, a fifth, orbiting, observatory, the Evolved Laser Interferometer Space Antenna, or e-LISA, is on the cards. The first pieces of apparatus designed to test the idea of e-LISA are already in space and the first LISA pathfinder results are very encouraging.

Together, by jointly forming a telescope that will permit astronomers to pinpoint whence the waves come, these devices will open a new vista onto the universe. (On the science side, the data is analyzed jointly by members of both LIGO and VIRGO, even though these data only come from LIGO. This is due to the analysis teams now being fully integrated. As this is not widely known, people do not realize that there is a large contribution from VIRGO scientists to the observations and to the future.) As technology improves, waves of lower frequency—corresponding to events involving larger masses—will become detectable. Eventually, astronomers should be able to peer at the first 380,000 years after the Big Bang, an epoch of history that remains inaccessible to every other kind of telescope yet designed.

The real prize, though, lies in proving Einstein wrong. For all its prescience, the theory of relativity is known to be incomplete because it is inconsistent with the other great 20th-century theory of physics, quantum mechanics. Many physicists suspect that it is in places where conditions are most extreme—the very places that launch gravitational waves—that the first chinks in relativity's armor will be found, and with them we will get a glimpse of a more all-embracing theory.

Gravitational waves, of which Einstein remained so uncertain, have provided direct evidence for black holes, about which he was long uncomfortable, and may yet yield a peek at the Big Bang, an event he knew his theory was inadequate to describe. They may now lead to his theory's unseating. If so, its epitaph will be that in predicting gravitational waves, it predicted the means of its own demise.

**Acknowledgments:** GFS acknowledges Laboratoire APC-PCCP, Université Paris Diderot, Sorbonne Paris Cité(DXCACHEXGS), and also the financial support of the UnivEarthS Labex program at Sorbonne Paris Cité (ANR-10-LABX-0023 and ANR-11-IDEX-0005-02). JLCC acknowledges financial support from conacyt Project 269652 and Fronteras Project 281.

**Author Contributions:** JLCC, SGU, and GFS conceived the idea and equally contributed.
**Conflicts of Interest:** The authors declare no conflict of interest.